\documentclass[prl,twocolumn,superscriptaddress]{revtex4}

\usepackage{graphicx,enumerate}
\usepackage{amsmath}
\usepackage{subfigure}

\newcommand{\mathd}{\mathrm{d}}
\newcommand{\tmem}[1]{{\em #1\/}}

\newcommand{\afm}{antiferromagnetic Ising model}

\newenvironment{enumeratenumeric}{\begin{enumerate}[1.] }{\end{enumerate}}

\begin{document}

\title{Beyond Cahn-Hilliard-Cook: Early Time Behavior of Symmetry
  Breaking Phase Transition Kinetics}

\author{Kipton Barros}
\email{kbarros@bu.edu}
\affiliation{Department of Physics, Boston University, Boston, Massachusetts 02215}
\affiliation{Center for Computational Science, Boston University, Boston, Massachusetts 02215}

\author{Rachele Dominguez}
\affiliation{Department of Physics, Boston University, Boston, Massachusetts 02215}

\author{W. Klein}
\affiliation{Department of Physics, Boston University, Boston, Massachusetts 02215}
\affiliation{Center for Computational Science, Boston University, Boston, Massachusetts 02215}

\begin{abstract}
  We extend the early time ordering theory of Cahn, Hilliard, and Cook
  (CHC) so that our generalized theory applies to solid-to-solid
  transitions. Our theory involves spatial {\tmem{symmetry breaking}}
  (the initial phase contains a symmetry not present in the final
  phase). The predictions of our generalization differ from those of
  the CHC theory in two important ways: exponential growth does not
  begin immediately following the quench, and the objects that grow
  exponentially are not necessarily Fourier modes. Our theory is
  consistent with simulation results for the long-range
  antiferromagnetic Ising model.
\end{abstract}

\pacs{64.70.K-, 64.60.De, 64.60.-i}

\maketitle

The early time dynamics of systems quenched into unstable states is of
considerable interest~
\cite{cahn_free_1959,cook_brownian_1970,langer_new_1975,
  gunton_dynamics_1983,binder_nucleation_1984,elder_role_1989,
  hyde_modelling_1996, mainville_x-ray_1997}. The first effective theory to treat this
process was developed by Cahn and Hilliard~\cite{cahn_free_1959} and
by Cook~\cite{cook_brownian_1970}. The CHC theory applies to processes
such as spinodal decomposition and continuous
ordering~\cite{gunton_dynamics_1983} and predicts that the early
evolution of the equal time structure factor following the quench is
characterized by exponentially growing Fourier modes. A primary
assumption of the CHC theory is that the initial configuration
following the quench is an unstable stationary point of the free
energy~\cite{corberi_early_1995}. This assumption fails for
solid-to-solid transitions.

In this paper we introduce a generalized theory which describes the
early time kinetics of spatial symmetry breaking transitions. We will
show that the kinetics can be separated into two well defined stages
for systems with effective long-range interactions. In the first stage
symmetry breaking fluctuations grow non-exponentially. In the second
stage the evolution crosses over to exponential growth analogous to
CHC. When the initial phase is a solid, we predict that the objects
which grow exponentially are not Fourier modes.

Binder~\cite{binder_nucleation_1984} has shown that the CHC theory is
valid only when the effective interaction range is large, $R \gg
1$~\cite{foot1},
as has been confirmed in Ising model simulations
~\cite{marro_time_1975, heermann_test_1984}. There is evidence that
many physical systems, such as polymers~\cite{binder_nucleation_1984}
and metals~\cite{cherne_non-classical_2004} have effective long-range
interactions. It is therefore natural to develop our theory in the
context of a long-range model. If the order parameter is expanded in
powers of $R^{-d/2}$~\cite{grant_theory_1985}, we can separate the
\emph{background} (of order $R^0=1$) from the noise induced
\emph{fluctuations} (of order $R^{-d/2}$). We will see that the CHC
theory describes evolving fluctuations with a fixed background.

Our early-time theory applies when the background is
non-stationary. Such a background occurs for example in the early-time
kinetics of a solid-to-solid transition where the lattice quickly
contracts or expands before before significant symmetry change
occurs. We will show that the evolution of the background is described
by noiseless dynamics which explicitly preserves the rotational and
translational symmetries of the initial phase. We say that the
transition involves spatial \emph{symmetry breaking} if the initial
phase contains symmetries not present in the final phase. When spatial
symmetry breaking occurs, we will show that the background evolves to
a stationary phase that is unstable with respect to symmetry breaking
fluctuations. Our theory predicts that the growth of these
fluctuations changes from non-exponential (stage~1) to exponential
(stage~2) when the background converges. The growth of the symmetry
breaking fluctuations is described by a linear theory which is valid
for a time of order $\ln R$~ \cite{binder_nucleation_1984}. For $R$
sufficiently large, both stages of our theory are observed.

We develop our theory in the context of a time-dependent
Ginzburg-Landau model with explicit long-range Kac
interactions~\cite{kac_van_1963}. The non-conserved field
$\phi(\vec{x}, t)$ plays the role of an order parameter and evolves
according to the Langevin dynamics
\begin{equation}
\label{dynamics1} \frac{\partial \phi(\vec{x}, t)}{\partial t} = -M \frac{\delta F_R [\phi]}{\delta \phi(\vec{x}, t)} + \sqrt{M} \tilde{\eta}(\vec{x}, t).
\end{equation}
$F_R [\phi]$ is the free energy of the configuration $\phi$ at time
$t$, and $R$ represents the effective interaction range. The
Gaussian white noise $\tilde{\eta}(\vec{x}, t)$ has zero mean and
second moment
$\left\langle \tilde{\eta}(\vec{x}, t) \tilde{\eta}(\vec{x}', t')
\right\rangle = k_B T \delta(t - t') \delta(\vec{x} - \vec{x}')$.
We set $M = 1$ corresponding to the rescaling of time,
$t \rightarrow t' = t/M$. The drift term is given by
\begin{subequations}
\begin{align}
-\frac{\delta F_R [\phi]}{\delta \phi(\vec{x})} & = \!\int d^d \vec{x}' \Lambda_R(\vec{x}') \phi(\vec{x} - \vec{x}') + f(\phi(\vec{x})) + h \label{free energy}\\
& = (\Lambda_R \ast \phi)(\vec{x}) + f(\phi(\vec{x})) + h,
\end{align}
\end{subequations}
where $\Lambda_R$ is a Kac potential of the form
$\Lambda_R(\vec{x}) = R^{-d} \Lambda(\vec{x}/R)$. The function $f$
represents entropic forces, and $h$ is an external field or chemical
potential. By convention, $f (\phi) |_{\phi = 0} = 0$.

We scale all lengths by $R$ so that Eq.~\eqref{dynamics1} simplifies
to
\begin{equation}
\label{dynamics2}
\frac{\partial u(\vec{r}, t)}{\partial t} = -\frac{\delta F [u]}{\delta u(\vec{r}, t)} + R^{-d/2} \eta(\vec{r}, t),
\end{equation}
where $\vec{r} = \vec{x}/R$, and
\begin{align}
u(\vec{r}, t) & = \phi(\vec{x}, t) \label{scaled_vars}\\
- \frac{\delta F [u]}{\delta u(\vec{r})} & = (\Lambda \ast u)(\vec{r}) + f(u(\vec{r})) + h.
\end{align}
The parameter $R$ in Eq.~\eqref{dynamics1} appears solely as
a prefactor to the noise term. The term
$\eta(\vec{r}, t) = R^{d/2} \tilde{\eta}(\vec{x}, t)$ represents
Gaussian white noise with zero mean and second moment
$\left\langle \eta( \vec{r}, t) \eta(\vec{r}', t') \right\rangle = k_B
T \delta(t - t') \delta (\vec{r} - \vec{r}')$,
which follows from the identity
$a^{-d} \delta( \vec{x}/a) = \delta(\vec{x})$.

The form of Eq.~\eqref{dynamics2} suggests expanding $u$ in the small
parameter $R^{-d/2}$:
\begin{equation}
\label{u exp} u = u^{(0)} + R^{-d/2} u^{(1)} + R^{-d} u^{(2)} + \ldots
\end{equation}
We substitute Eq.~\eqref{u exp} into Eq.~\eqref{dynamics2} and obtain the
dynamical equations
\begin{align}
\frac{\partial u^{(0)}}{\partial t} & = - \frac{\delta F [u^{(0)}]}{\delta u} = \Lambda \ast u^{(0)} + f(u^{(0)}) + h \label{hierarchy}\\
\frac{\partial u^{(1)}}{\partial t} & = \mathcal{L} u^{(1)} + \eta, \label{hierarchy2}
\end{align}
where
\begin{equation}
\label{L} \mathcal{\mathcal{L}} \psi = \Lambda \ast \psi + f'(u^{(0)}) \psi,
\end{equation}
and $f'(u) = \mathd f/\mathd u$. We remark that the nonlinear dynamics
of $u^{(0)}$ in Eq. (\ref{hierarchy}) is deterministic and decoupled from higher
orders. The dynamics of $u^{(1)}$ is stochastic, linear, and depends
on $u^{(0)}$ through $\mathcal{L}$.

As we have mentioned, the CHC theory emerges as the evolution of
$u^{(1)}$ when $u^{(0)}$ is a stationary point of the free energy.
Let us see how this works for a disorder-order transition occurring
after a rapid quench from infinite to finite temperature and $h = 0$
(recall that $f(0) = 0$). At $t = 0$ the system is initially
disordered so Eq.~\eqref{hierarchy} has the trivial solution
$u^{(0)} = 0$ for all time. With this solution, Eq.~\eqref{hierarchy2}
can be solved in Fourier space,
\begin{align}
u^{(1)}(\vec{k}, t) & = u^{(1)}(\vec{k}, 0) e^{D(\vec{k}) t} \nonumber \\
& \quad + \!\int_0^t dt'\! e^{D(\vec{k})(t - t')} \eta(\vec{k}, t'), \label{cosolution}
\end{align}
where $D(\vec{k}) = \Lambda(\vec{k}) + f'(u^{(0)} = 0)$. The structure
factor $S(k, t) = \left\langle |\phi|^2 \right\rangle/V$ can be
calculated using Eq.~\eqref{scaled_vars}, thus reproducing the CHC
theory. For spin systems the volume $V$ equals the total number of
spins because the lattice spacing is taken to be unity.

We can easily determine the time scale for which the CHC theory is
applicable. Equation~\eqref{u exp} is meaningful when the neglected
$O(R^{-d})$ terms are small. One requirement is that
$R^{-d/2} u^{(1)} \ll u^{(0)} \simeq 1$. The exponential growth of
$u^{(1)}$ from Eq.~\eqref{cosolution} suggests that the linear theory
breaks down at a time $t \sim \ln
R$~\cite{binder_nucleation_1984,heermann_test_1984}.

For many phase transitions (such as solid-to-solid) we need to
consider the evolution of both $u^{(0)}$ and
$u^{(1)}$. Equation~\eqref{hierarchy2} predicts exponential growth of
$u^{(1)}(t)$ whenever $\mathcal{L}$ is time independent, which from
Eq.~\eqref{hierarchy} occurs when $\delta F/\delta u^{(0)}(x, t) = 0$.
In general, the initial condition $u^{(0)}(t=0)$ will not be such a
stationary point. We will show that, due to symmetry breaking,
$u^{(0)}$ converges to an unstable stationary configuration
$u^{\ast}$. Correspondingly, $\mathcal{L}$ will converge to a time
independent operator. This instability of $u^{\ast}$ means that
$\mathcal{L}$ will have positive eigenvalues, corresponding to
the unstable symmetry breaking growth modes.

Let $G_i$ and $G_f$ represent the symmetry groups of rotations and
translations of the initial and final phases respectively. We will
now show that $u^{(0)}(\vec{r}, t)$ is invariant under $G_i$ provided
that the potential $\Lambda(\vec{r})$ shares the rotational symmetries
of $G_i$. Discretization of Eq.~\eqref{hierarchy} with the time step
$\Delta t$ yields
\begin{equation}
\label{discretized} u^{(0)}_{t + \Delta t} = u^{(0)}_t + \Delta t \big(\Lambda \ast u_t^{(0)} + f(u_t^{(0)}) + h \big).
\end{equation}
If $u^{(0)}$ is invariant under $G_i$ at time $t$, it is invariant at
time $t + \Delta t$, which follows from the properties of the
convolution operation $(\ast)$ and the rotational symmetries of
$\Lambda$. Induction establishes the $G_i$ symmetry of $u^{(0)}$ for
all $t$~\cite{meaningful}.

How does $u^{(0)}$ evolve for a phase transition with symmetry
breaking? We see from Eq.~\eqref{hierarchy} that $F [u^{(0)}]$ is
non-increasing. Physically, $F$ must be bounded from below, so we
expect $u^{(0)}$ to converge to some configuration $u^{\ast}$. This
convergence occurs on a time scale independent of $R$. Symmetry
considerations ensure that $u^{\ast}$ is not the stable phase: there
exists a spatial transformation $g$ which is in $G_i$ but not in $G_f$
(the symmetry breaking condition) under which the configuration
$u^{\ast}$ and not the stable phase is invariant. Because $u^{\ast}$
is not the stable phase, we expect that $u^{\ast}$ is an unstable free
energy stationary point. Simultaneous to the evolution of $u^{(0)}$,
the dynamical noise induces symmetry breaking fluctuations in
$u^{(1)}$ which are of magnitude $R^{-d/2}$. These fluctuations are
unstable, and if $u^{(0)}$ has converged, will grow exponentially for
a time proportional to $\ln R$, analogous to the predictions of
CHC.

\begin{figure}
\subfigure[]{
\includegraphics[scale=0.55]{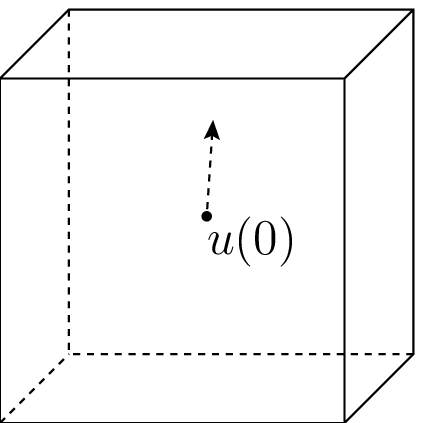}} \quad
\subfigure[]{
\includegraphics[scale=0.55]{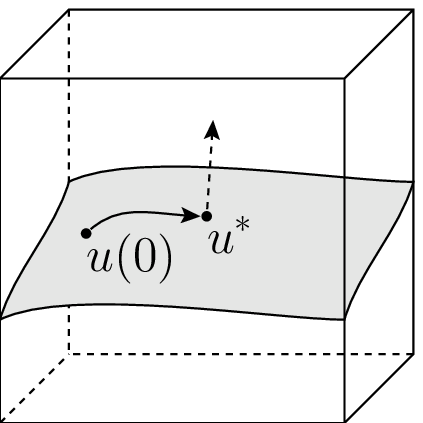}}
\quad \subfigure[]{
\includegraphics[scale=0.55]{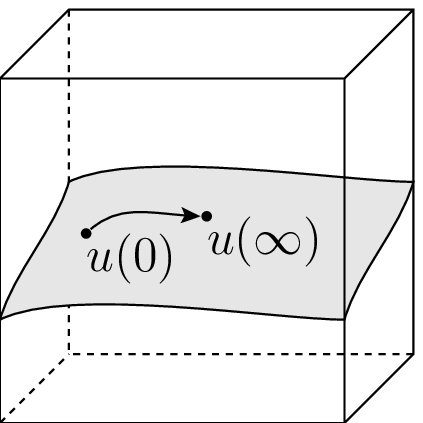}}
\caption{\label{breaking} (a) The CHC theory is applicable if the
  initial configuration $u(0)$ is a free energy stationary point. CHC
  describes the immediate exponential growth of Fourier modes, lasting
  a time $t \sim \ln R$. (b) For symmetry breaking transitions (e.g.,
  solid-to-solid) the early time kinetics of $u(t)$ has two stages. In
  the first stage the leading order contribution to $u$ evolves
  deterministically and non-linearly to a symmetry-constrained (shaded
  plane) free energy minimum $u^{\ast}$ in a time $t \sim 1$. In the
  second stage, symmetry breaking modes grow exponentially for a time
  $t \sim \ln R$. (c) Without symmetry breaking (e.g., solid-to-fluid)
  the leading order contribution to $u$ evolves deterministically to
  the stable phase $u(\infty)$ in a time $t \sim 1$.}
\end{figure}

We conclude that spatial symmetry breaking phase transition kinetics can
be decomposed into two stages:
\begin{enumeratenumeric}
\item $t \lesssim t_0$: Nonlinear evolution of $u^{(0)}$ toward
  $u^{\ast}$, a configuration of minimum free energy subject to
  symmetry constraints. The configuration $u^{\ast}$ is not the stable
  phase. The dynamical equation for $u^{(1)}$ is linear but has an
  explicit time dependence. Note that $t_0$ is independent of $R$.

\item $t_0 \lesssim t \lesssim \ln R$: To a good approximation
  $u^{(0)}$ has converged to $u^{\ast}$. The linear theory of
  $u^{(1)}$ becomes analogous to the CHC theory, and describes
  exponential growth of the unstable symmetry breaking modes.
\end{enumeratenumeric}
These two stages are illustrated in Fig.~\ref{breaking}(b). In
contrast, there is no stage~1 process in the CHC theory, as
illustrated in Fig.~\ref{breaking}(a).

Phase transition kinetics {\tmem{without}} spatial
symmetry breaking, such as solid-to-fluid, are qualitatively
different. Here $u^{(0)}$ will evolve to $u^{\ast}$ but, unlike the
symmetry breaking case, $u^{\ast}$ is the stable
phase because the symmetries of $G_i$ are included in $G_f$ (no
symmetries are broken in the transition). Note that all the interesting
dynamics in this transition occurs through $u^{(0)}$, which is
independent of the noise. This process is illustrated in
Fig.~\ref{breaking}(c).

Let us see how exponential growth arises in the second stage of a
symmetry breaking transition by considering
Eq.~\eqref{hierarchy}. Because $\mathcal{L}$ is a real and symmetric
linear operator, it has a complete orthonormal eigenbasis and real
eigenvalues~\cite{footL}. The
eigenvectors of $\mathcal{L}$ are Fourier modes only if $u^{(0)}$ is
uniform. We can express the dynamics of $u^{(1)}$ in the eigenbasis of
$\mathcal{L}$:
\begin{equation}
\label{eigendynamics} \frac{\partial u_v^{(1)}}{\partial t}= \sum_{v'} \mathcal{L}_{vv'} u_{v'}^{(1)} + \eta_v = \lambda_v u_v^{(1)} + \eta_v,
\end{equation}
where $v$ and $\lambda_v$ represent the corresponding eigenvectors and
eigenvalues of $\mathcal{L}$. The subscripts indicate eigenbasis
components, for example
$u_v = \!\int d^d \vec{r} v(\vec{r}) u(\vec{r})$ and
$\mathcal{L}_{vv'} = \!\int d^d \vec{r} v(\vec{r}) \mathcal{L}
v'(\vec{r}) = \lambda_v \delta_{vv'}$.
The eigenvectors are normalized, and we can show that
$\left\langle \eta_v(t) \eta_{v'}(t') \right\rangle = \delta_{vv'}
\delta(t - t')$.

For times $t \gtrsim t_0$ the operator $\mathcal{L}$ is time
independent and Eq.~\eqref{eigendynamics} can be solved directly:
\begin{equation}
\label{eigen sol} u_v^{(1)}(t) = u_v^{(1)}(t_0) e^{\lambda_v(t - t_0)} + \!\int_{t_0}^t dt' e^{\lambda_v(t - t')} \eta_v(t').
\end{equation}
The exponential growth of $u^{(1)}$ is apparent. We can express
$u^{(1)}$ in the Fourier basis,
\begin{equation}
u^{(1)}(\vec{k}, t) = \sum_v v(\vec{k}) u^{(1)}_v(t), \end{equation}
where $v(\vec{k})$ is the Fourier representation of the eigenvector $v$. If
$R$ is sufficiently large and there is a single largest eigenvalue
$\lambda_v$, then a single eigenvector $v$ will grow exponentially faster than
all others. In this case, and at sufficiently large times, we can approximate
\begin{equation}
u^{(1)}(\vec{k}, t) \approx v(\vec{k}) \Big[ u_v^{(1)}(t_0) e^{\lambda_v(t - t_0)} + \!\int_{t_0}^t\! dt' e^{\lambda_v(t - t')} \eta_v(t') \Big].
\end{equation}
We see that the exponential growth of the eigenvector $v$ implies
exponential growth of all the Fourier modes of $u^{(1)}$,
$\left\langle |u^{(1)}(\vec{k}, t) |^2 \right\rangle \propto e^{2 \lambda_v t}$.
These Fourier modes eventually dominate all other contributions to the structure factor
$S = \left\langle | \phi(\vec{k}, t) | \right\rangle/V$, provided that
the linear theory is valid ($t \lesssim \ln R$).

\begin{figure}
\includegraphics[scale=0.85]{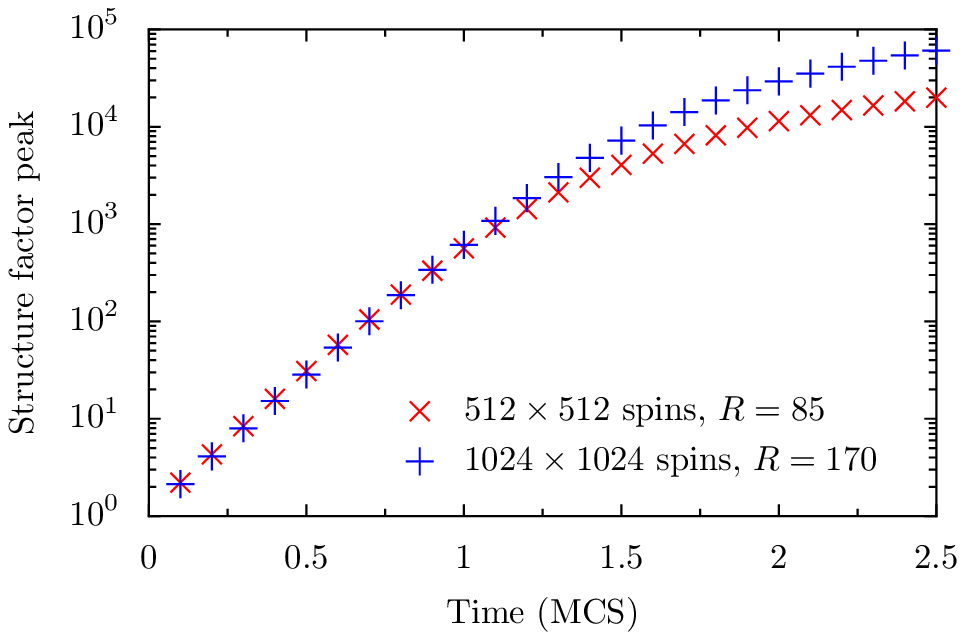}
\caption{\label{critical quench}Evolution of the structure factor peak
  ($\left\langle | \phi_{\max} |^2 \right\rangle/V$) for the
  fluid-solid transition in the \afm\ following a critical ($h = 0$)
  quench. The CHC theory correctly predicts exponential growth,
  beginning immediately after the quench for this transition.}
\end{figure}

\begin{figure}
\includegraphics[scale=0.85]{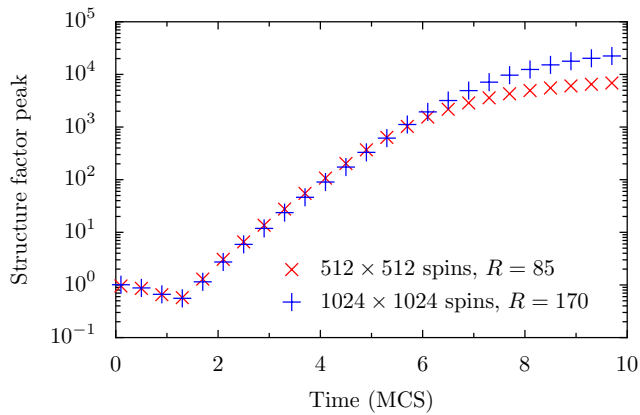}
\caption{\label{off critical quench}Evolution of the structure factor
  peak ($\left\langle |\phi_{\max} |^2 \right\rangle/V$) for the
  fluid-solid transition following an off-critical ($h \neq 0$)
  quench. The initial growth has two stages, confirming the prediction
  of our generalized theory. The first stage is non-exponential and is
  independent of $R$. The second stage is exponential growth of the
  symmetry breaking modes (in this case Fourier modes) in analogy to
  the CHC theory.}
\end{figure}

We now compare our generalized theory to simulations of the 2D
antiferromagnetic Ising model with a long-range square
interaction. This model contains a disordered \tmem{fluid} phase, as
well as \tmem{clump}, and \tmem{stripe} solid
phases~\cite{dominguez_early_2008,similar}. In the clump phase,
localized regions of enhanced magnetization are arranged on a square
lattice. In the stripe phase, regions of enhanced magnetization are
arranged in periodic stripes. All fluid-to-solid phase transitions
involve symmetry breaking, as do the transitions between clump and
stripe phases. In contrast, solid-to-fluid transitions do not involve
symmetry breaking because the uniform fluid phase contains all
possible spatial symmetries.

For the Ising model with $R \gg 1$ a free energy functional $F[\phi]$
can be derived~\cite{dominguez_early_2008}. Rather than the dynamics
of Eq.~\eqref{dynamics1} we use single-spin flip Monte Carlo dynamics
to simulate the system. At each update a spin is selected at random
and flipped with the Glauber transition probability
$p = (1 + e^{\beta \Delta E})^{-1}$. Time is measured in units of
Monte Carlo steps per spin (MCS).

In Figs.~\ref{critical quench} and \ref{off critical quench} we
display the peak of the structure factor,
$S (\vec{k}, t) = \left\langle \phi (\vec{k}, t)^2 \right\rangle/V$,
for fluid-to-solid phase transitions following critical ($h = 0$) and
off-critical ($h = 0.8$) quenches. In both cases the temperature is
reduced from $T = \infty$ to 0.05. The critical and off-critical
transitions are described, respectively, by the CHC theory and our
generalization. As predicted, the off-critical dynamics can be
separated into two stages: initial non-exponential growth followed by
an extended period of exponential growth. The growth modes are Fourier
modes for both types of quenches considered because the initial phase
is disordered.

In summary, we have shown that the CHC theory can be generalized to
describe solid-to-solid transitions. The key ingredient of this
generalization is spatial symmetry breaking. The predictions of our
generalized theory differ from those of the CHC theory in two
fundamental ways: (1) the exponential growth of the symmetry breaking
modes does not immediately follow the quench, and (2) these symmetry
breaking modes are not generally Fourier modes. We have performed
simulations of the long-range \afm\ for the off-critical
fluid-to-solid transition, and have confirmed the existence of a
transient stage preceding exponential growth of the structure
factor. A future paper will show simulations confirming this theory in
symmetry breaking transitions~\cite{dominguez_early_2008}. Finally, we
point out that our theory does not apply in the presence of symmetry
breaking defects in the initial conditions.

\begin{acknowledgments}
  We thank Harvey Gould, Louis Colonna-Romano, and Minghai Li for
  useful discussions. This material is based upon work supported by
  NSF Grant No.\ DGE-0221680 (K.B.) and DOE Grant No. 2234-5 (K.B.,
  R.D., W.K.)
\end{acknowledgments}


\begin{thebibliography}{17}
\expandafter\ifx\csname natexlab\endcsname\relax\def\natexlab#1{#1}\fi
\expandafter\ifx\csname bibnamefont\endcsname\relax
\def\bibnamefont#1{#1}\fi
\expandafter\ifx\csname bibfnamefont\endcsname\relax
\def\bibfnamefont#1{#1}\fi
\expandafter\ifx\csname citenamefont\endcsname\relax
\def\citenamefont#1{#1}\fi
\expandafter\ifx\csname url\endcsname\relax
\def\url#1{\texttt{#1}}\fi
\expandafter\ifx\csname urlprefix\endcsname\relax\def\urlprefix{URL }\fi
\providecommand{\bibinfo}[2]{#2}
\providecommand{\eprint}[2][]{\url{#2}}

\bibitem[{\citenamefont{Cahn and Hilliard}(1959)}]{cahn_free_1959}
\bibinfo{author}{\bibfnamefont{J.~W.} \bibnamefont{Cahn}} \bibnamefont{and}
\bibinfo{author}{\bibfnamefont{J.~E.} \bibnamefont{Hilliard}},
\bibinfo{journal}{J. Chem. Phys.} \textbf{\bibinfo{volume}{31}},
\bibinfo{pages}{688} (\bibinfo{year}{1959}).

\bibitem[{\citenamefont{Cook}(1970)}]{cook_brownian_1970}
\bibinfo{author}{\bibfnamefont{H.~E.} \bibnamefont{Cook}}, \bibinfo{journal}{Act. Metall.} \textbf{\bibinfo{volume}{18}}, \bibinfo{pages}{297} (\bibinfo{year}{1970}).

\bibitem[{\citenamefont{Langer et~al.}(1975)\citenamefont{Langer, Bar-On, and Miller}}]{langer_new_1975}
\bibinfo{author}{\bibfnamefont{J.}~\bibnamefont{Langer}}, \bibinfo{author}{\bibfnamefont{M.}~\bibnamefont{Bar-On}}, \bibnamefont{and} \bibinfo{author}{\bibfnamefont{H.}~\bibnamefont{Miller}}, \bibinfo{journal}{Phys. Rev. A} \textbf{\bibinfo{volume}{11}}, \bibinfo{pages}{1417} (\bibinfo{year}{1975}).

\bibitem[{\citenamefont{Gunton et~al.}(1983)\citenamefont{Gunton, Miguel, and Sahni}}]{gunton_dynamics_1983}
\bibinfo{author}{\bibfnamefont{J.~D.} \bibnamefont{Gunton}}, \bibinfo{author}{\bibfnamefont{M.~S.} \bibnamefont{Miguel}}, \bibnamefont{and} \bibinfo{author}{\bibfnamefont{P.}~\bibnamefont{Sahni}}, in \emph{\bibinfo{booktitle}{Phase Transitions and Critical Phenomena}}, edited by \bibinfo{editor}{\bibfnamefont{C.}~\bibnamefont{Domb}} \bibnamefont{and} \bibinfo{editor}{\bibfnamefont{J.~L.} \bibnamefont{Lebowitz}} (\bibinfo{publisher}{Academic Press}, \bibinfo{address}{London}, \bibinfo{year}{1983}), Vol.~\bibinfo{volume}{8}.

\bibitem[{\citenamefont{Binder}(1984)}]{binder_nucleation_1984}
\bibinfo{author}{\bibfnamefont{K.}~\bibnamefont{Binder}}, \bibinfo{journal}{Phys. Rev. A} \textbf{\bibinfo{volume}{29}}, \bibinfo{pages}{341} (\bibinfo{year}{1984}).

\bibitem[{\citenamefont{Elder and Desai}(1989)}]{elder_role_1989}
\bibinfo{author}{\bibfnamefont{K.~R.} \bibnamefont{Elder}} \bibnamefont{and} \bibinfo{author}{\bibfnamefont{R.~C.} \bibnamefont{Desai}}, \bibinfo{journal}{Phys. Rev. B} \textbf{\bibinfo{volume}{40}}, \bibinfo{pages}{243} (\bibinfo{year}{1989}).

\bibitem[{\citenamefont{Mainville et~al.}(1997)\citenamefont{Mainville, Yang, Elder, Sutton, Ludwig, and Stephenson}}]{mainville_x-ray_1997}
\bibinfo{author}{\bibfnamefont{J.}~\bibnamefont{Mainville}}, \bibinfo{author}{\bibfnamefont{Y.~S.} \bibnamefont{Yang}}, \bibinfo{author}{\bibfnamefont{K.~R.} \bibnamefont{Elder}}, \bibinfo{author}{\bibfnamefont{M.}~\bibnamefont{Sutton}}, \bibinfo{author}{\bibfnamefont{J.}~\bibnamefont{Ludwig}}, \bibnamefont{and} \bibinfo{author}{\bibfnamefont{G.~B.} \bibnamefont{Stephenson}}, \bibinfo{journal}{Phys. Rev. Lett.} \textbf{\bibinfo{volume}{78}}, \bibinfo{pages}{2787} (\bibinfo{year}{1997}).

\bibitem[{\citenamefont{Hyde et~al.}(1996)\citenamefont{Hyde, Sutton, Harris, Cerezo, and Gardiner}}]{hyde_modelling_1996}
\bibinfo{author}{\bibfnamefont{J.~M.} \bibnamefont{Hyde}}, \bibinfo{author}{\bibfnamefont{A.~P.} \bibnamefont{Sutton}}, \bibinfo{author}{\bibfnamefont{J.~R.~G.} \bibnamefont{Harris}}, \bibinfo{author}{\bibfnamefont{A.}~\bibnamefont{Cerezo}}, \bibnamefont{and} \bibinfo{author}{\bibfnamefont{A.}~\bibnamefont{Gardiner}}, \bibinfo{journal}{Modell. Simul. Mater. Sci. Eng.} \textbf{\bibinfo{volume}{4}}, \bibinfo{pages}{33} (\bibinfo{year}{1996}).


\bibitem[{\citenamefont{Corberi et~al.}(1995)\citenamefont{Corberi, Coniglio, and Zannetti}}]{corberi_early_1995}
\bibinfo{author}{\bibfnamefont{F.}~\bibnamefont{Corberi}}, \bibinfo{author}{\bibfnamefont{A.}~\bibnamefont{Coniglio}}, \bibnamefont{and} \bibinfo{author}{\bibfnamefont{M.}~\bibnamefont{Zannetti}}, \bibinfo{journal}{Phys. Rev. E} \textbf{\bibinfo{volume}{51}}, \bibinfo{pages}{5469} (\bibinfo{year}{1995}).

\bibitem{foot1}The interaction range $R$ is measured in units of a fundamental length scale such as the lattice spacing or the intermolecular separation.

\bibitem[{\citenamefont{Marro et~al.}(1975)\citenamefont{Marro, Bortz, Kalos, and Lebowitz}}]{marro_time_1975}
\bibinfo{author}{\bibfnamefont{J.}~\bibnamefont{Marro}}, \bibinfo{author}{\bibfnamefont{A.~B.} \bibnamefont{Bortz}}, \bibinfo{author}{\bibfnamefont{M.~H.} \bibnamefont{Kalos}}, \bibnamefont{and} \bibinfo{author}{\bibfnamefont{J.~L.} \bibnamefont{Lebowitz}}, \bibinfo{journal}{Phys. Rev. B} \textbf{\bibinfo{volume}{12}}, \bibinfo{pages}{2000} (\bibinfo{year}{1975}).

\bibitem[{\citenamefont{Heermann}(1984)}]{heermann_test_1984}
\bibinfo{author}{\bibfnamefont{D.~W.} \bibnamefont{Heermann}}, \bibinfo{journal}{Phys. Rev. Lett.} \textbf{\bibinfo{volume}{52}}, \bibinfo{pages}{1126} (\bibinfo{year}{1984}).

\bibitem[{\citenamefont{Cherne et~al.}(2004)\citenamefont{Cherne, Baskes, Schwarz, Srinivasan, and Klein}}]{cherne_non-classical_2004}
\bibinfo{author}{\bibfnamefont{F.~J.} \bibnamefont{Cherne}}, \bibinfo{author}{\bibfnamefont{M.~I.} \bibnamefont{Baskes}}, \bibinfo{author}{\bibfnamefont{R.~B.} \bibnamefont{Schwarz}}, \bibinfo{author}{\bibfnamefont{S.~G.} \bibnamefont{Srinivasan}}, \bibnamefont{and} \bibinfo{author}{\bibfnamefont{W.}~\bibnamefont{Klein}}, \bibinfo{journal}{Modell. Simul. Mater. Sci. Eng.} \textbf{\bibinfo{volume}{12}}, \bibinfo{pages}{1063} (\bibinfo{year}{2004}).

\bibitem[{\citenamefont{Grant et~al.}(1985)\citenamefont{Grant, Miguel, Vials, and Gunton}}]{grant_theory_1985}
\bibinfo{author}{\bibfnamefont{M.}~\bibnamefont{Grant}}, \bibinfo{author}{\bibfnamefont{M.~S.} \bibnamefont{Miguel}}, \bibinfo{author}{\bibfnamefont{J.}~\bibnamefont{Vials}}, \bibnamefont{and} \bibinfo{author}{\bibfnamefont{J.~D.} \bibnamefont{Gunton}}, \bibinfo{journal}{Phys. Rev. B} \textbf{\bibinfo{volume}{31}}, \bibinfo{pages}{3027} (\bibinfo{year}{1985}).

\bibitem[{\citenamefont{Kac et~al.}(1963)\citenamefont{Kac, Uhlenbeck, and Hemmer}}]{kac_van_1963}
\bibinfo{author}{\bibfnamefont{M.}~\bibnamefont{Kac}}, \bibinfo{author}{\bibfnamefont{G.~E.} \bibnamefont{Uhlenbeck}}, \bibnamefont{and} \bibinfo{author}{\bibfnamefont{P.~C.} \bibnamefont{Hemmer}}, \bibinfo{journal}{J. Math. Phys.} \textbf{\bibinfo{volume}{4}}, \bibinfo{pages}{216} (\bibinfo{year}{1963}).

\bibitem{meaningful}Note, however, that $u^{(0)}$ is a meaningful approximation to $u$ only when the $R^{-d/2}$ expansion is valid, i.e., $t \lesssim \ln R$.

\bibitem{footL}We assume that $\mathcal{L}$ is a finite matrix, corresponding to a finite system volume discretized in space.

\bibitem[{\citenamefont{Dominguez et~al.}()\citenamefont{Dominguez, Barros, and Klein}}]{dominguez_early_2008}
\bibinfo{author}{\bibfnamefont{R.}~\bibnamefont{Dominguez}}, \bibinfo{author}{\bibfnamefont{K.}~\bibnamefont{Barros}}, \bibnamefont{and} \bibinfo{author}{\bibfnamefont{W.}~\bibnamefont{Klein}}, \bibinfo{note}{in preparation}.

\bibitem{similar}Similar structures have been observed in other models with repulsive interactions. See, for example, \cite{glaser_soft_2007}.

\bibitem[{\citenamefont{Glaser et~al.}(2007)\citenamefont{Glaser, Grason, Kamien, Kosmrlj, Santangelo, and Ziherl}}]{glaser_soft_2007}
\bibinfo{author}{\bibfnamefont{M.}~\bibnamefont{Glaser}}, \bibinfo{author}{\bibfnamefont{G.~M.} \bibnamefont{Grason}}, \bibinfo{author}{\bibfnamefont{R.~D.} \bibnamefont{Kamien}}, \bibinfo{author}{\bibfnamefont{A.}~\bibnamefont{Kosmrlj}}, \bibinfo{author}{\bibfnamefont{C.~D.} \bibnamefont{Santangelo}}, \bibnamefont{and} \bibinfo{author}{\bibfnamefont{P.}~\bibnamefont{Ziherl}}, \bibinfo{journal}{Europhys. Lett.} \textbf{\bibinfo{volume}{78}}, \bibinfo{pages}{46004} (\bibinfo{year}{2007}).

\end{thebibliography}
\end{document}